\documentclass[aps,prc,twocolumn,showpacs,preprintnumbers,superscriptaddress,
amsmath,amssymb]{revtex4}
\usepackage{graphicx}% Include figure files
\usepackage{dcolumn}% Align table columns on decimal point
\usepackage{bm}% bold math

\begin{document}
\def\MOSCOW{Institute for Nuclear Research, Russian Academy of Sciences,
117312 Moscow, Russia}
\def\Mainz{Institut f{\"u}r Kernphysik, Johannes Gutenberg-University, Mainz,
D-55099 Germany}
\newcommand{\goo}{\,\raisebox{-.5ex}{$\stackrel{>}{\scriptstyle\sim}$}\,}
\newcommand{\loo}{\,\raisebox{-.5ex}{$\stackrel{<}{\scriptstyle\sim}$}\,}

\title{Production of hypernuclei in multifragmentation of nuclear
spectator matter}

\author{A.S.~Botvina}       \affiliation{\MOSCOW}
\author{J.~Pochodzalla}      \affiliation{\Mainz}

\date{\today}

\begin{abstract}
In peripheral collisions of relativistic heavy ions highly excited
spectators containing $\Lambda$ hyperons can be produced. Such
strange spectator matter may undergo a break-up into many fragments
(multifragmentation) as it is well established for ordinary nuclear
systems. We generalize the statistical multifragmentation model,
previously successfully used for the description of experimental
data, for the case of hypernuclear systems. We predict relative
yields of hypernuclei and the main characteristics of such a
break-up. We point at a connection of this phenomenon with a
liquid-gas phase transition in hypermatter.
\end{abstract}


\maketitle

One promising way to study hypernuclei is to use the copious
production of hyperons in relativistic heavy ion collisions. As seen
in experiments in the GeV range \cite{Lopez}, $\Lambda$ hyperons
originating from the hot participant region have a broad rapidity
distribution and a considerable fraction can be found in the
spectator kinematical region. In peripheral collisions these
$\Lambda$ hyperons may be captured by excited spectators.
Subsequently these spectators may decay producing cold and possibly
exotic hypernuclei. It has been demonstrated experimentally that the
cross section of this reaction may be on the order of a few
microbarns \cite{Dubna}. A similar value can be obtained with UrQMD
calculations by assuming a coalescence of the produced $\Lambda$
hyperons with spectators \cite{hyphi}. Indeed there are extensive
plans to investigate hypernuclei produced in such heavy ion
reactions using stable \cite{Dubna} as well as exotic beams
\cite{hyphi}.

An important feature of these reactions is a correlation between the
excitation energy transferred to the spectator matter and the number
of high energy particles produced as a result of collisions between
nucleons in the participant region (see e.g. \cite{Turzo}), which
are also related to the strangeness production. As a consequence the
excited spectators break-up into many fragments (multifragmentation)
\cite{SMM,Pochodzalla1,EOS}, and the captured $\Lambda$ will be
finally attached to these fragments. According to the present
understanding, multifragmentation is a relatively fast process, with
a characteristic time around 100 fm/c, where, nevertheless, a high
degree of equilibration (chemical equilibrium) is reached
\cite{EPJrev}. This is a result of the strong interaction between
the nucleons, and we believe that a small admixture of other
particles like hyperons interacting with similar cross sections
\cite{pdg} will not change the general statistical behaviour.
Multifragmentation can also be associated with the liquid-gas phase
transition in nuclear systems, and thermodynamical features of this
transition were experimentally investigated \cite{Pochodzalla}. As
was demonstrated by numerous comparisons of theory and experiment,
statistical models allow a successful description of experimental
data \cite{SMM,EOS,Botvina95,Xi}. In this letter we demonstrate that
multifragmentation of hypermatter can be described in a similar way,
and some properties of hypernuclei can be explored by examining
relative yields of hyperfragments. The calculations also show that
the usage of exotic heavy ion beams will indeed allow to produce
hypernuclei with extreme N/Z ratios and thus hypernuclei which are
not experimentally accessible with any other technique.

For the theoretical description of this process we modify the
grand-canonical version of the Statistical Multifragmentation Model
(SMM) \cite{SMM,SJNP85}. Previously, this version, with explicit
addition of mass, charge, energy and momentum conservations, was
successfully used for the description of multifragmentation in
relativistic heavy-ion reactions \cite{EOS,Botvina95,Xi}.
Microcanonical extensions of statistical models, like
\cite{Botvina01}, are essential only for a few details of the phase
transition, but not for the general description of the produced
fragments. The model assumes that a hot nuclear spectator with total
mass (baryon) number $A_0$, charge $Z_0$, number of $\Lambda$
hyperons $H_0$, and temperature $T$ expands to a low density
freeze-out volume, where the system is in chemical equilibrium. The
statistical ensemble includes all break-up channels composed of
nucleons and excited fragments. The primary fragments are formed in
the freeze-out volume $V$. We use the excluded volume approximation
$V=V_0+V_f$, where $V_0=A_0/\rho_0$ ($\rho_0\approx$0.15 fm$^{-3}$
is the normal nuclear density), and parametrize the free volume
$V_f=\kappa V_0$, with $\kappa \approx 2$
%, as taken in refs.
\cite{EOS,Botvina95,Xi}.

Nuclear clusters in the freeze-out volume are described as follows:
light fragments with mass number $A < 4$ are treated as elementary
particles with corresponding spin and translational degrees of
freedom ("nuclear gas"). Their binding energies were taken from
experimental data \cite{SMM,Bando,Hashimoto}. Fragments with $A=4$
are also treated as gas particles with table masses, however, some
excitation energy is allowed $E_{x}=AT^{2}/\varepsilon_0$
($\varepsilon_0 \approx$16 MeV is the inverse volume level density
parameter \cite{SMM}), that reflects the presence of excited states
in $^{4}$He, $^{4}_{\Lambda}$H, and $^{4}_{\Lambda}$He nuclei.
Fragments with $A > 4$
%($Z$ is their charge, and $H$ is the number of $\Lambda$)
are treated as heated liquid drops. In this way one can study the
nuclear liquid-gas coexistence of hypermatter in the freeze-out
volume. The internal free energies of these fragments are
parametrized as the sum of the bulk ($F_{A}^B$), the surface
($F_{A}^S$), the symmetry ($F_{AZH}^{\rm sym}$), the Coulomb
($F_{AZ}^C$), and the hyper energy ($F_{AH}^{\rm hyp}$):
\begin{equation}
F_{AZH}(T,V)=F_{A}^B+F_{A}^S+F_{AZH}^{\rm sym}+F_{AZ}^C+F_{AH}^{\rm hyp}~~.
\end{equation}
Here, $H$ denotes the number of $\Lambda$'s. The first three terms
are written in the standard liquid-drop form \cite{SMM}:
%\begin{eqnarray}
%\begin{equation*}
%$F_{A}^B(T)=\left(-w_0-\frac{T^2}{\varepsilon_0}\right)A$,
$F_{A}^B(T)=(-w_0-\frac{T^2}{\varepsilon_0})A$,~
%\end{equation}
%\begin{equation}
$F_{A}^S(T)=\beta_0\left(\frac{T_c^2-T^2}{T_c^2+T^2}\right)^{5/4}A^{2/3}$,
and
%\end{equation}
%\begin{equation}
$F_{AZH}^{\rm sym}=\gamma \cdot{(A-H-2Z)^2}/{(A-H)}$.
%\end{equation*}
%\end{eqnarray}
The model parameters $w_0=16$ MeV, $\beta_0=18$ MeV, $T_c=18$ MeV
and $\gamma=25$ MeV were extracted from nuclear phenomenology and
provide a good description of multifragmentation data
\cite{SMM,EOS,Botvina95,Xi}. The Coulomb interaction of the
fragments is described within the Wigner-Seitz approximation, and
$F_{AZ}^C$ is taken as in ref.~\cite{SMM}.
%:
%\begin{equation}
%F_{AZ}^C(V)=\frac{3}{5}\left[1-\left(\frac{V_0}{V}\right)^{1/3}\right]
%\frac{(eZ)^2}{r_0A^{1/3}}~~.
%\end{equation}
%where $r_0=1.2$ fm and $e$ denotes the electron charge.

The new term is the free hyper-energy $F_{AH}^{\rm hyp}$. We assume
that it does not change with temperature, i.e., it is determined
solely by the binding energy of the hyper-fragments. Presently, only
few ten masses of single hypernuclei (mostly light ones) are
experimentally established \cite{Bando,Hashimoto}, and there is only
very limited information on double hypernuclei available. However,
there are some theoretical estimations of their masses based on a
description of the available data. One of them is the Samanta
formula \cite{Samanta} which suggests a hyper term
\begin{equation}
E_{sam}^{\rm hyp}=H\cdot(-10.68+48.7/(A^{2/3})).
\label{eq_samanta}
\end{equation}
The contribution proportional to $A^{-2/3}$ is motivated by
calculations of the $\Lambda$ binding energy in a potential well
\cite{Rote}. In what later follows we use $F_{AH}^{\rm
hyp}=E_{sam}^{\rm hyp}$ as one of the versions for our calculations.
We have also explored another hyper term (in the following we call
it the liquid drop hyper term):
\begin{equation}
F_{AH}^{\rm hyp}=(H/A)\cdot(-10.68 A + 21.27 A^{2/3}).
\label{eq_ldrop}
\end{equation}
In this formula the binding energy is proportional to the fraction
of hyperons in the system ($H/A$). The second part represents the
volume contribution reduced by the surface term and thus resembles a
liquid-drop parametrization based on the saturation of the nuclear
interaction. The linear dependence at small $H/A$ is in agreement
with theoretical predictions \cite{Greiner} for hyper matter. We
have found that for single hypernuclei with $A>8$ the average
deviation of the liquid-drop masses from the experimental masses
tabulated in \cite{Bando} is 1.8 MeV, while using the Samanta formula
this deviation is worse and amounts to
2.2 MeV. If not mentioned otherwise we use the liquid-drop hyper
term in the following.

At this point we should note that the properties of the primary
fragments  may change in the medium compared to the vacuum due to
the proximity of other fragments. Presently, there are evidences
that the symmetry energy \cite{LeFevre,Iglio,Souliotis} and surface
energy \cite{Botvina06} of hot fragments in multifragmentation may
be modified. However, these corrections are not large and influence
multifragmentation in finite nuclear systems very little. They are
more important for astrophysical applications \cite{Botvina05}.
Moreover, since the binding energies of hypernuclei are mostly not
known even in vacuum, this problem is naturally included in
searching for a reliable mass formula for hypernuclei within this
approach.

% fig.1 --------------------------------------------------------
\begin{figure}[t]
\includegraphics[width=0.8\linewidth]{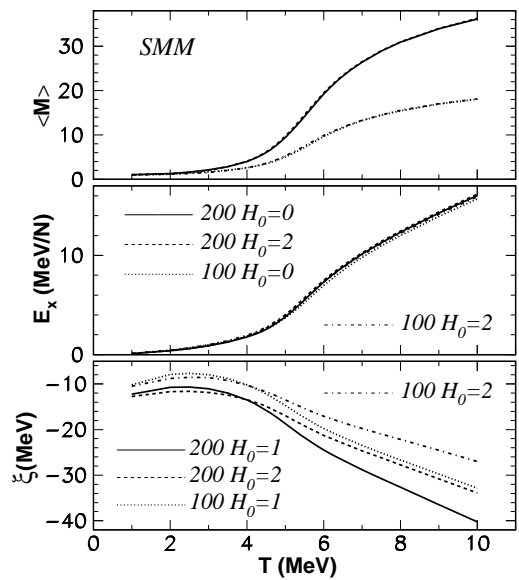}
\caption{{SMM calculations of multifragmentation of spectator
sources with $A_0$=200, $Z_0$=80 and $A_0$=100, $Z_0$=40, with
number of $\Lambda$ hyperons $H_0$=0, 1, 2. Top panel -- average
multiplicity of produced primary fragments (notations are as in the
middle panel), middle panel -- excitation energy of the sources
(caloric curve), bottom panel -- chemical potential $\xi$
responsible for strangeness, as function of temperature $T$.}}
\label{fig:botvina01}
\end{figure}
% --------------------------------------------------------------
The break-up channels should be selected according to their
statistical weight. In the Grand Canonics this leads to the following
average yields of individual fragments:
\begin{eqnarray} \label{yazh}
Y_{\rm AZH}=g_{\rm AZH}V_f\frac{A^{3/2}}{\lambda_T^3}
{\rm exp}\left[-\frac{1}{T}\left(F_{AZH}-\mu_{AZH}\right)\right],
\nonumber\\
\mu_{AZH}=A\mu+Z\nu+H\xi~,
\end{eqnarray}
Here $g_{\rm AZH}$ is the ground-state degeneracy factor of species
$(A,Z,H)$, $\lambda_T=\left(2\pi\hbar^2/m_NT\right)^{1/2}$ is the
nucleon thermal wavelength, $m_N \approx$939 MeV is the average
nucleon mass. In our case $H<<A$, therefore, the mass difference
between nucleons and $\Lambda$ can be disregarded in this
expression. The chemical potentials $\mu$, $\nu$, and $\xi$ are
responsible for the mass (baryon) number, charge, and strangeness
conservation in the system. They can be found from the balance
equations:
\begin{eqnarray}
\sum_{AZH}A Y_{\rm AZH}=A_0,
\sum_{AZH}Z Y_{\rm AZH}=Z_0,
\sum_{AZH}H Y_{\rm AZH}=H_0.
\nonumber
\end{eqnarray}

%% fig.2 --------------------------------------------------------
\begin{figure}[t]
\includegraphics[width=0.75\linewidth]{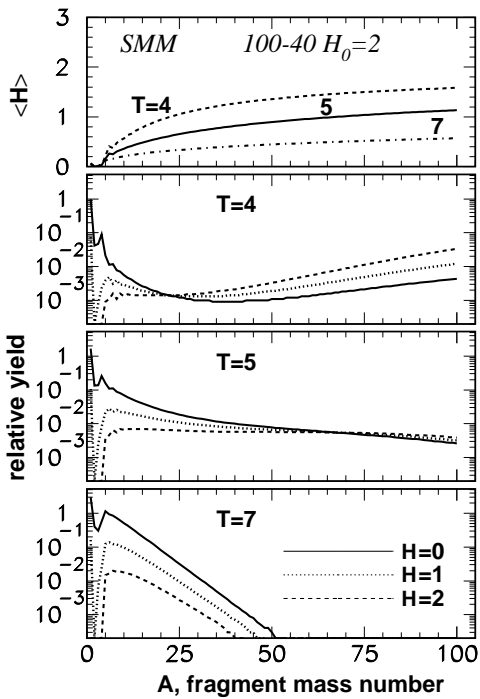}
\caption{{Multifragmentation of sources with $A_0$=100, $Z_0$=40,
$H_0$=2. Top panel -- average number of hyperons in fragments at
temperatures $T$=4, 5, and 7 MeV. Three bottom panels -- relative
yields of fragments (per event), separately for fragments containing
$H$=0, 1, or 2 $\Lambda$, at the same temperatures.}}
\label{fig:botvina02}
\end{figure}
% --------------------------------------------------------------
Within this extended SMM approach we have performed calculations for
excited spectator sources which can be produced during peripheral
relativistic heavy ion collisions. Below, we show results for heavy
systems with $A_0$=200 and $Z_0$=80, and intermediate systems with
$A_0$=100 and $A_0$=50, varying the charge-to-mass ratio ($Z/A$)
from 0.4 to 0.5, and $H_0$ from 0 to 2. The first source has the
mass and $Z/A$ ratio typical for a heavy projectile. We expect the
most extensive absorption of $\Lambda$ hyperons by spectators in the
case of heavy projectiles and targets. It will be possible to
increase the strangeness production by selecting smaller impact
parameters, leading to a higher excitation and a smaller mass of the
projectile spectators. However, previous analyses \cite{LeFevre}
suggest that their $Z/A$ ratio will be nearly conserved.
%% fig.3 --------------------------------------------------------
\begin{figure}[t]
\includegraphics[width=0.80\linewidth]{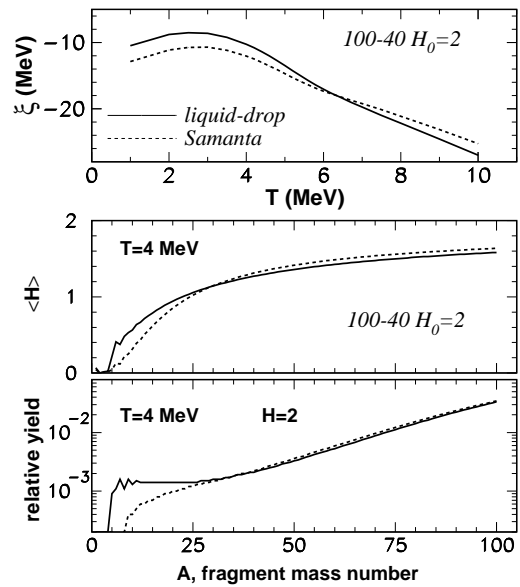}
\caption{{Comparison of SMM calculations with the liquid-drop and
Samanta descriptions of hyper terms in the mass formula, for the
same sources as in Fig.~2. Top panel -- the strangeness chemical
potential $\xi$ versus temperature $T$. Middle panel -- average
number of $\Lambda$ hyperons in fragments, and bottom panel --
yields of fragments with two $\Lambda$, at $T$=4 MeV.}}
\label{fig:botvina03}
\end{figure}
% ---------------------------------------------------------------
%% fig.4 --------------------------------------------------------
\begin{figure}[t]
\includegraphics[width=0.75\linewidth]{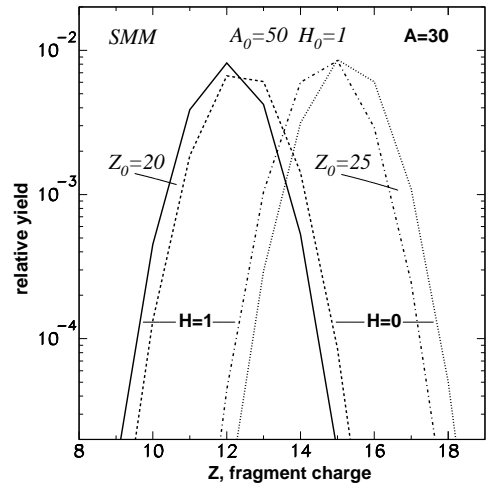}
\caption{{Charge distributions of primary fragments with $A$=30
produced after break-up of sources with $A_0$=50, $H_0$=1, $Z_0$=20
(solid and dashed lines), and $Z_0$=25 (dotted and dot-dashed
lines), at temperature $T$=5 MeV. $H=0$ stands for fragments without
$\Lambda$, $H=1$ are fragments with one $\Lambda$.}}
\label{fig:botvina04}
\end{figure}
% --------------------------------------------------------------

In Fig.~\ref{fig:botvina01} we present the average multiplicity of
fragments produced after the break-up of the excited spectators, the
caloric curve for these break-ups, and the chemical potential $\xi$,
versus the temperature of the spectator matter. At small
temperatures ($T<3$ MeV), the compound nucleus channel ($M=1$)
dominates. In the following it should decay by means of evaporation
or fission processes \cite{SMM}. One can consider it as a nuclear
liquid phase. At temperatures $T=4-6$ MeV the system produces mainly
several intermediate mass fragments (IMF, with $4<A\loo$ $A_0$/3),
however, a heavy residue and light particles are also possible. In
nuclear multifragmentation this region is interpreted as a
liquid-gas phase coexistence \cite{SMM,Pochodzalla}. Here the
caloric curve signals an increasing heat capacity, associated with a
rapid rise of the excitation energy. And at high excitations
($T>8-9$ MeV) the system disintegrates mainly into light particles
(a gas phase). One can see that the small admixture of $\Lambda$'s
does not change this general behaviour. However, a drop of $\xi$ at
$T>4$ MeV signals a decreasing average strangeness content of the
fragments.
%, since the latter are produced in a larger amount.

In the top panel of Fig.~\ref{fig:botvina02} we explicitly
demonstrate this decrease of the average number of $\Lambda$'s
contained in the fragments with rising temperature. In three bottom
panels we show how the mass distributions of fragments evolve with
the temperature. Around the onset of multifragmentation ($T\approx
4$ MeV) there is a % so-called
"U-shaped" distribution with a big
residue and few small fragments. The %big
residue contains nearly all
strangeness of the system, while the small fragments are mainly
without hyperons. Also the share of free $\Lambda$'s is not
essential, hyperons prefer to be bound in fragments in all cases.
When the system breaks into many fragments, the strangeness is
distributed among all produced IMF. Within the phase coexistence
region ($T\approx 5$ MeV) we have a rather flat distributions for
the produced big residues, and nearly equal probabilities for
producing the residues with different number of hyperons. Large
fluctuations of number of fragments and their size are expected in
this case.

We have found that there is a sensitivity of the fragment yields in
multifragmentation to the mass formulae used for description of
binding energy of hypernuclei. In Fig.~\ref{fig:botvina03} we
compare SMM calculations performed with the liquid-drop hyper term
(\ref{eq_ldrop}) in free energy of individual fragments, and with the Samanta
term (\ref{eq_samanta}). There is a clear difference in the chemical potential
$\xi$, and, as a result, the yields of hyper-fragments are also
different. As one can see from the bottom panels the liquid-drop
formula predicts more strangeness in IMF's than the Samanta formula.
The difference in the yields is particularly large for small double
hyper-fragments. In future, this observable may allow to test
experimentally different mass formulas for hypernuclei in
multifragmentation.

It is an important feature that the isotope composition of the
produced hypernuclei can be varied considerably in these reactions.
In Fig.~\ref{fig:botvina04} we show charge distributions of
fragments with A=30 after the break-up of spectator sources with
different isospin. The corresponding $Z_0/A_0$ ratios can be easily
obtained by selecting different projectiles. After
% As a result of
multifragmentation, the hot primary fragments have approximately the
same average isospin content as the sources, whereas the widths of
the charge distributions depend mainly on the fragment symmetry
energy \cite{Botvina01}.
%One can see in Fig.~\ref{fig:botvina04} that
Thus the present work shows that it is indeed possible to produce
very neutron rich hyper-fragments. We expect that the final isospin
distributions will be similar to the ones reported in previous
multifragmentation studies with exotic beams \cite{EPJrev}.

After the break-up, the Coulomb acceleration and the secondary
de-excitation of primary hot fragments should be taken into account
\cite{SMM}. As it is well known from calculations and experiments,
the de-excitation of nuclei with $A\leq 200$ will proceed mainly by
emission of nucleons.
%(and, a little bit, fission).
As a consequence the resulting distributions of cold nuclei are not
very different from the primary ones, they are just shifted to lower
masses. In this case the reported regularities will not change, and
the differences in relative yields will survive, though some of them
may become smaller \cite{SMM,Iglio,Souliotis}.
%In the
%forthcoming papers we'll include the secondary decay processes for
%both normal and hyper-fragments \cite{Botvina87,Alicia}, and perform
%calculations for the whole set of spectators produced at different
%impact parameters, after the dynamical stage of the heavy-ion
%collisions \cite{Botvina95,Xi}.
The last stage will be the mesonic or non-mesonic decay of
hypernuclei which can used for the identification of the
hypernuclei~\cite{hyphi}.

We conclude that multifragmentation reactions offer a new
possibility for investigating hypernuclei under conditions
essentially different from those accessible in conventional nuclear
structure studies. % at low energies.
One of the advantages is that in this reaction one can abundantly
produce hypernuclei with unusual $Z/A$ ratios. It is encouraging
that one can distinguish between different mass formulae of
hypernuclei, since their properties are manifested in their relative
yields. More generally, we suggest that this reaction can be
considered as an experimental tool to study clusterization of
nuclear matter with strangeness, and the phase diagram of
hyper-nuclear matter at densities $\rho \approx 0.1-0.3 \rho_0$ and
temperatures around $T \approx$ 3--8 MeV reached in the freeze-out
volume. Apparently, such a phase transition may take place in some
astrophysical cases, for example, in neutron star crusts.

The authors thank I.N. Mishustin, T.R. Saito and W. Trautmann for
stimulating discussions.
%A.S.B. thanks Institut f\"ur Kernphysik at
%the University Mainz for hospitality.
This work was partly supported
by the Bundesministerium f\"ur Bildung, Wissenschaft,
Forschung und Technologie, Germany, under contract number
06-MZ-225I.

\end{document}